\documentclass[a4paper]{article}
\usepackage{array,booktabs}
\usepackage{geometry}
\usepackage{amsmath,amssymb}
\usepackage{graphicx}
\usepackage{color,xcolor}
\usepackage{amsfonts,amssymb}
\usepackage{longtable}
\usepackage{subfigure}
\usepackage[english]{babel}
\usepackage[utf8]{inputenc}
\usepackage{amsmath}
\usepackage{graphicx}
\usepackage{animate}
\usepackage{natbib}
\usepackage{epstopdf}
\usepackage[colorinlistoftodos]{todonotes}
\usepackage{colortbl}
\usepackage{float}
\usepackage{amsfonts,amssymb}
\usepackage{multirow}
\graphicspath{{./frames/}}

\title{A Semi-Parametric Estimation Method for the Quantile Spectrum with an Application to Earthquake Classification Using Convolutional Neural Network}
\makeatletter
\let\@fnsymbol\@arabic
\makeatother
\author{Tianbo Chen$^1$; Ying Sun\thanks{King Abdullah University of Science and Technology (KAUST), Statistics Program, Thuwal 23955-6900, Saudi Arabia. E-mails: tianbo.chen@kaust.edu.sa; ying.sun@kaust.edu.sa.}\text{ }; Ta-Hsin Li\thanks{IBM Watson Research Center, NY, US. E-mail: thl@us.ibm.com.} }

\date{}

\begin{document}
\maketitle
\begin{abstract}
\noindent In this paper, a new estimation method is
introduced for the quantile spectrum, which uses a parametric form of the
autoregressive (AR) spectrum coupled with nonparametric smoothing. The method begins with quantile
periodograms which are constructed by trigonometric quantile regression
at different quantile levels, to represent the
serial dependence of time series at various quantiles. At each quantile level, we approximate the quantile spectrum by a function in the form of
an ordinary AR spectrum.  In this model, 
we first compute what we call the quantile autocovariance function (QACF) by the inverse Fourier
transformation of the quantile periodogram at each quantile level. Then, we solve the Yule-Walker equations formed by the QACF to obtain the quantile partial
autocorrelation function (QPACF) and the scale parameter.   Finally, we smooth
QPACF and the scale parameter across the quantile levels using a nonparametric smoother, convert the
smoothed QPACF to AR coefficients, and obtain the AR spectral density function. Numerical
results show that the proposed method outperforms
other conventional smoothing techniques. We take advantage of the two-dimensional property of the estimators and train a convolutional neural
network (CNN) to classify smoothed quantile periodogram of earthquake
data and achieve a higher  accuracy than a similar classifier using ordinary periodograms.
\end{abstract}

\section{Introduction}
Motivated by the least-squares interpretation of the ordinary periodogram, \cite{li2012quantile} proposed the quantile periodogram based on trigonometric quantile regression. Similarly to the behavior of the ordinary spectrum and periodogram \citep{brockwell1984time}, the quantile periodogram has an asymptotic exponential distribution but the mean function, called the quantile spectrum, is a scaled version of the ordinary spectrum of the level-crossing process. Related works include \citet{hagemann2013robust}, \cite{li2016time}, \citet{dette2015copulas} \cite{lim2016composite}, \citet{birr2017quantile,birr2019model}, and \citet{barunik2019quantile}.

The quantile periodogram was inspired by the quantile regression \citep{koenker2001quantile}, which examines the conditional quantiles and, thus, provides a richer view of the data than the traditional least-squares regression.  Just as the ordinary periodogram can be  obtained by applying least squares regression to the trigonometric regressor, the quantile periodogram is computed by applying quantile regression instead of least squares regression. Since quantile regression is known for its robustness against outliers  and heavy-tailed noise, quantile periodograms are also endowed with these properties. \cite{li2012detection} used the quantile periodogram to detect and estimate hidden periodicity from noisy observations when the noise distribution was asymmetric with a heavy tail on one side, while ordinary periodograms are less effective in handling such noise. \citet{li2014quantile} applied the quantile periodogram to Dow Jones Industrial Average and showed the advantages of quantile periodograms when handling time-dependent variance.

In traditional spectral analysis, the ordinary periodogram needs to be smoothed to serve as an estimator of the underlying ordinary (or power) spectrum. Many methods have been proposed to do so. One classical method is to apply a nonparametric smoother to the periodogram across frequencies. \cite{shumway2016time} discussed several periodogram smoothing techniques include moving-average smoothing. \cite{wahba1980automatic}  developed the optimally smoothed spline (OSS) estimator, and the smoothing parameter is selected to minimize the expected integrated mean square error. The span selection is an important issue in periodogram smoothing. The unbiased risk estimator is used to produce the span selector in \cite{lee1997simple}, and the selector does not require strong conditions on the spectral density function. Another popular method for spectral density estimation is based on likelihood. \cite{capon1983maximum} used the maximum-likelihood filter to produce the minimum-variance unbiased estimator of the spectral density function. \cite{chow1985sieve} proposed a sieve for the estimation of the spectral density of a Gaussian stationary stochastic process using likelihood, and compared to  standard periodogram-based estimate, the proposed method aims at exploiting the full Gaussian nature of the process. The well-known Whittle likelihood was developed for time series analysis in \cite{whittle1953estimation, whittle1954some}, and an approximation to the likelihood is constructed from the spectrum and periodogram.  In \cite{pawitan1994nonparametric}, the spectral density function is estimated by the penalized Whittle likelihood. For parametric estimation methods, the autoregressive (AR) spectral approximation is discussed in \cite{shumway2016time}. \cite{chan1982spectral} and \citet{friedlander1984modified} used the Yule-Walker method to estimate the parameters.  

Similar to the ordinary periodogram, the quantile periodogram is also very noisy at each quantile level. To estimate the copula rank-based quantile spectrum, \citet{zhang2019bayesian} produced an automatically smoothed estimator for  the copula spectral density kernel (CSDK), along with samples from the posterior distributions of the parameters via a Hamiltonian Monte Carlo (HMC) step. \citet{hagemann2013robust} applied the \citet{parzen1957consistent} class of kernel spectral density estimators to the quantile periodogram and characterized the asymptotic properties of the quantile and smoothed quantile periodograms. \citet{dette2015copulas} showed that the quantile spectrum can be estimated consistently by a window smoother of the quantile periodogram.

The ordinary spectrum is a critical feature of Gaussian time series, and it is often used to classify or cluster time series. For example, \citet{orhan2011eeg}  introduced a classifier based on multilayer perceptron neural network (MLPNN) as a diagnostic-decision support mechanism in epilepsy treatment. \citet{maadooliat2018nonparametric} employed ordinary spectral density functions for electroencephalogram (EEG) data clustering. The quantile spectrum contains information at both the frequency levels and the quantile levels. Due to the additional information it provides compared to the ordinary periodogram, the quantile periodogram is expected to produce better classification and clustering results. The two-dimensional (2D) nature of the quantile spectrum offers opportunities to apply image-based techniques. The convolutional neural network (CNN) is an effective and popular technique for 2D image classification. For example, \citet{krizhevsky2012imagenet} trained a large, deep CNN to classify 1.3 million high-resolution images in the LSVRC-2010 ImageNet training set. For accurate classification, a good estimation of the quantile spectrum is crucial. The quality of the estimator directly affects the classification performance.

Motivated by the function approximation capability of the autoregressive (AR) spectrum \citep{shumway2011spectral}, we propose a new method that uses the parametric form of the AR spectrum to approximate the quantile spectrum at each quantile level. The novelty of our approach is threefold. First, at each quantile level, we approximate the quantile spectrum by the ordinary AR spectral density derived from what we call the quantile autocovariance function as the Fourier transform of the quantile periodogram, where the order of the AR model is automatically selected by the AIC. Furthermore,  we apply nonparametric smoothing techniques with respect to the quantile levels for the AR orders as well as the AR coefficients, where the latter is carried out in the partial autocorrelation domain to ensure the causality of the approximating AR spectrum. Therefore, the proposed method ensures not only the smoothness with respect to the frequencies but also the smoothness with respect to the quantile levels. Finally, we treat the smoothed quantile periodograms as images and apply image classification techniques to classify time series. For the case study, we employ the proposed quantile spectral estimation technique, coupled with a CNN classifier, to classify a set of time series for earthquakes. The results show that the proposed quantile periodogram estimator is a better choice than the ordinary periodogram for the purpose of time series classification.

The rest of the paper is organized as follows. In Section \ref{q2}, we introduce the quantile periodogram and the proposed estimation procedure. The simulation study is performed in Section \ref{q3}, and the time series classification of earthquake waveforms using both ordinary and quantile periodograms is described in Section \ref{q4}. We conclude and discuss the paper in Section \ref{q5}.

\section{Methodology}\label{q2}
In this section, we introduce the quantile spectrum and quantile periodogram (Section \ref{q21}). We describe the Yule-Walker approach (Section \ref{q22}) and the AR model representation of the quantile periodogram (Section \ref{q23}).
\subsection{Quantile Spectrum and Quantile Periodogram} \label{q21}
Let $Y_t$ be a stationary process, and let $F(u)$ denote the marginal cumulative distribution function (CDF) of $\{Y_t\}$. Assume $F(u)$ is a continuous function with strictly positive derivative $\dot{F}(u)$, the level-crossing process of  $Y_t$ is defined as $\{I(Y_t\le \lambda_{\tau})\}$, where $\lambda_{\tau}$ is the $\tau$-quantile of $Y_t$: $E\{I(Y_t\le \lambda_{\tau})\}=\tau$. The level-crossing spectrum is defined as:
$$
f^*_{\tau}(\omega)=\sum_{k=-\infty}^{\infty}\{1-\frac{1}{2\tau(1-\tau)}S_k \}{\rm exp}(2\pi j\omega k),
$$
where $S_k={\rm Pr}\{(Y_{t+k}-\lambda_{\tau})( Y_t- \lambda_{\tau})<0\}$ is the lag-$k$ level-crossing rate and $\omega \in (0,1/2)$. The quantile spectrum \citep{li2012quantile} is defined as
\begin{equation}\label{eq1}
f_{\tau}(\omega)=\eta_{\tau}^2 f^*_{\tau}(\omega),
\end{equation}
where $\eta_{\tau}=\tau(1-\tau)^{1/2}/\dot{F}(\lambda_{\tau})$ is the scaling factor determined by the marginal distribution.

Given observations $Y_1,...,Y_n$, consider the quantile regression problem
$$
\hat{\boldsymbol{\beta}}_{n,\tau}(\omega):={\sf argmin}_{\boldsymbol{\beta} \in \mathbb{R}^2,\lambda_{\tau} \in \mathbb{R}}\sum_{t=1}^n\rho_{\tau} \{ Y_t-\lambda_{\tau}-{\bf x}_t^{\top}(\omega)\boldsymbol{\beta}(\omega)\},
$$
where
$\rho_{\tau}(u)=u\{\tau-I(u<0)\}$, ${\bf x}_t(\omega)=\{{\sf cos}(2\pi \omega t),{\sf sin}(2 \pi\omega t)\}^{\top}$ for $\omega \in (0,1/2)$.
The quantile periodogram of $\{Y_1,...,Y_n\}$ at quantile level $\tau$ is defined as
$$
Q_{n,\tau}(\omega):=\frac{1}{4}n||\hat{\boldsymbol{\beta}}_{n,\tau}(\omega)||_2^2=\frac{1}{4}n\hat{\boldsymbol{\beta}}^{\top}_{n,\tau}(\omega) \hat{\boldsymbol{\beta}}_{n,\tau}(\omega).
$$
In other words, the quantile periodogram is a scaled version of the squared norm (or sum of squares) of the quantile regression coefficients corresponding to the trigonometric regressor ${\bf x}_t(\omega)$. The Laplace periodogram \citep{li2008laplace} is a special case of the quantile periodogram with $\tau=0.5$. Note that the quantile periodogram can be generalized to quantile cross-periodograms by using two quantile levels $\tau_1 \ne \tau_2$ \citep{dette2015copulas,kley2016quantile,birr2017quantile}.

The quantile periodograms can be regarded
as generalizations of the ordinary periodogram
$$
I_n(\omega):=\frac{1}{n}\left|\sum_{t=1}^nY_te^{-2\pi jt\omega}\right|^2.
$$
It is easy to show that when $\omega$ is a Fourier frequency of the form $l/n$, with $l$ being integer, we can write
$I_n(\omega):=\frac{1}{4}n||\bar{\boldsymbol{\beta}}_n(\omega)||_2^2$,
where $\bar{\boldsymbol{\beta}}_n(\omega)$ is given by
$$\bar{\boldsymbol{\beta}}_n(\omega):={\sf argmin}_{\boldsymbol{\beta} \in \mathbb{R}^2,\mu \in \mathbb{R}}\sum_{t=1}^n \{ Y_t-\mu-{\bf x}_t^{\top}(\omega)\boldsymbol{\beta}(\omega)\}^2,$$ and $\mu$ is the sample mean. \citet{li2012quantile} proved that the quantile periodogram is asymptotically exponentially distributed with the quantile spectrum in (1) as the mean of the asymptotic exponential distribution.

\subsection{Yule-Walker Equations}\label{q22}
The Yule-Walker equations provide a way to obtain the autoregressive (AR) coefficients and the partial autocorrelation function (PACF) from the ordinary autocovariance function (ACF).  Given the ACF $r(h)$,  the AR coefficients $\phi(j)\text{ },j=1,...,p$ can be obtained from
\[
\left[\begin{array}{c}
r(1)\\
r(2) \\
\vdots\\
r(p-1)\\
r(p)
\end{array}\right]_{\bf r} =
\left[\begin{array}{ccccc}
r(0) & r(1)&  \cdots & r(p-1)\\
r(1) & r(0) &  \cdots & r(p-2)\\
\vdots &   &   \vdots \\
r(p-2)& r(p-3) &  \cdots & r(1) \\
r(p-1) & r(p-2) & \cdots & r(0)
\end{array}\right]_{\bf R}\left[\begin{array}{c}
\phi(1)\\
\phi(2) \\
\vdots\\
\phi(p-1)\\
\phi(p)
\end{array}\right]_{\boldsymbol{\phi}},
\] i.e.,  ${\bf r}={\bf R}\boldsymbol{\phi}$. Since {\bf R} is a full-rank symmetric Toeplitz matrix, the invertability is guaranteed and $\boldsymbol{\phi}$={\bf R}$^{-1}${\bf r}. Let {\bf r}$_{i}$ and {\bf R}$_{i}$ denote the case when $p=i$. The PACF $\psi$ and the residual variance $\sigma_p^2$ are obtained as follows:
\begin{itemize}
	\item set initial value $\sigma_{0}^2=r(0)$.
	\item Loop on $i$, $1\leq i \leq p$. Then,\\
	*\text{ } \text{ }$\boldsymbol{\phi}_{i}=$({\bf R}$_{i}$)$^{-1}${\bf r}$_i=[\phi_i(1),\phi_i(2),...,\phi_i(i)]^{\top}$;\\
	*\text{ } \text{ }$\psi(i)=\phi_i(i)$ and $\sigma_i^2=\sigma_{i-1}^2\{1-\phi_i^2(i)\}$;
	\item End loop on $i$. Then, we have the $\psi= \{\psi(1),...,\psi(p)\}$ and the residual variance $\sigma_p^2$.
\end{itemize}
The Levinson-Durbin algorithm \citep{brockwell2006time}, allows for a recursive implementation that solves the Yule-Walker equations efficiently.
\subsection{Autoregressive Approximation of Quantile Spectrum}\label{q23}

\subsubsection{Quantile Autocovariance Function and Order Selection}
Suppose that $h(\omega)$ is the ordinary spectrum. Then, given $\epsilon>0$, there exists an AR spectrum $f(\omega)$ such that $\left|f(\omega)-h(\omega) \right|<\epsilon$ for all $\omega \in (0, 1/2)$ \citep{shumway2016time}. Motivated by this function approximation capability, we use the ordinary AR spectrum of order $p_{\tau_i}$ to approximate the quantile spectrum at level $\tau_i$, which is denoted by $f_{\tau_i}$, for $i = 1,...,m$, where the AR order $p_{\tau_i}$ is quantile level specific. Suppose that $Q_{\tau_i}(\omega_l),\text{ }l=0,1,...,n-1$, is the raw quantile periodogram obtained from the data, where $\omega_l=l/n$ is the Fourier frequency in $[0,1)$. Here, the raw quantile periodogram is calculated at Fourier frequencies $\omega_l \in (0,1/2)$, and is extended to $[0,1)$ by symmetry (the quantile periodogram at $\omega_l=0$ is set to be 0). We define the quantile autocovariance function (QACF) by the inverse Fourier transform of $Q_{\tau_i}(\omega_l)$ as
\begin{equation}
\hat{\gamma}_{\tau_i}(h)=n^{-1}\sum_{l=0}^{n-1}e^{2\pi j \omega_l h}Q_{\tau_i}(\omega_l), h=0,1,...,n-1.
\end{equation}
By solving the Yule-Walker Equations in Section \ref{q22} using the Levinson-Durbin algorithm, we obtain what we call the quantile PACF (QPACF) $\psi_{\tau_i}(1),...,\psi_{\tau_i}(p_{\tau_i})$, the AR coefficients $\phi_{\tau_i}(1),...,\phi_{\tau_i}(p_{\tau_i})$ and the residual variance $\hat{\sigma}^{2}_{p_{\tau_i}}$ which we call the scale parameters. The resulting AR spectrum in (3) maximizes the entropy  $\int_0^1 log\{f(\omega)\} d\omega$ among all
spectral density functions whose first $p_{\tau_i}$ partial autocovariances coincide with the QACF in (2).

We propose to choose the order of the AR model by minimizing the Akaike information criterion (AIC) \citep{Akaike1974}, i.e.,
$$
\hat{p}_{\tau_i}={\rm argmin}_{p_{\tau_i}}\{n {\rm log}(\hat{\sigma}^2_{p_{\tau_i}})+2p_{\tau_i}\}.
$$
The AIC balances the goodness of fit with the complexity of the AR model at each quantile level. The bias-variance tradeoff is also shown in Bayesian information criterion (BIC) \citep{schwarz1978estimating}, corrected AIC (AIC$_c$) \citep{hurvich1989regression}, and finite sample information criterion (FSIC) \citep{broersen1998autoregressive}. Alternatively, we can use the same order $p$ for all quantile levels, with $p$ determined by minimizing the average AIC across the quantile levels.

There may also be a need to smooth the order sequence $\{\hat{p}_{\tau_1},...,\hat{p}_{\tau_m}\}$ and scale sequence $\{\hat{\sigma}^{2}_{p_{\tau_1}},...,\hat{\sigma}^{2}_{p_{\tau_m}}\}$ in order to reduce their variability across the quantile levels. To obtain  smooth order sequence and scale sequence, we apply Friedman's SuperSmoother \citep{friedman1984variable} because of its flexibility. Other nonparametric smoothers can also be applied. The smoothed versions are denoted by $\{\hat{p}^*_{\tau_1},...,\hat{p}^*_{\tau_m}\}$ and $\{\hat{\sigma}^{*2}_{p_{\tau_1}},...,\hat{\sigma}^{*2}_{p_{\tau_m}}\}$.

\subsubsection{Quantile PACF and AR Coefficients}
With $\hat{p}^*_{\tau_i},\text{ }i=1,..,m$ identified, we obtain the QPACF 
$$\hat{\Psi}_{\tau_i}=[\psi_{\tau_i}(1),...,\psi_{\tau_i}(\hat{p}^*_{\tau_i}),0,...,0],$$ 
in which we pad zeros at the end to make the length of the QPACFs equal to $\hat{p}^*={\rm max}_{i\in1,...,m}\{\hat{p}^*_{\tau_i}\}$. Then, we have the QPACF matrix:
$$
\hat{\Psi}=[\hat{\Psi}_{\tau_i},...,\hat{\Psi}_{\tau_m}]^{\top}.$$
We propose to smooth the columns of $\hat{\Psi}$ using a nonparametric smoother such as splines with the tuning parameters selected by the ordinary level-one-out cross-validation or generalized cross-validation (function {\sf smooth.spline} in {\sf R}). The smoothed quantile PACF matrix is denoted by $\hat{\Psi}^{*}.$

Then, the corresponding matrix
$\hat{\boldsymbol{\Phi}}^*$ of the quantile AR coefficients can be computed at each quantile level $\tau_i$ via the method described in \citet{barndorff1973parametrization}.
Let $\{ \psi^*_{\tau_i}(1),..., \psi^*_{\tau_i}(\hat{p}^*) \}$ be the $i$-th row of $\hat{\Psi}^*$. Then, 
\begin{itemize}
\item $\boldsymbol{\phi}_{\tau_i}^*=\psi^*_{\tau_i}(1)$

\item Loop on $k$, $2\leq k \leq \hat{p}^*$. Then,\\
$\boldsymbol{\phi}'=\boldsymbol{\hat{\phi}}_{\tau_i}^*$;\\
$\boldsymbol{\hat{\phi}}_{\tau_i}^*=\boldsymbol{\phi}'-\psi^*_{\tau_i}(k)\cdot{\sf rev}(\boldsymbol{\phi}')$, where ${\sf rev}(\cdot)$ reverses the order of the argument;\\
$\boldsymbol{\hat{\phi}}_{\tau_i}^*$={\sf append}$\{\boldsymbol{\hat{\phi}}^*_{\tau_i},\psi^*_{\tau_i}(k)\}$.
\item End loop on $k$. Then, return $\boldsymbol{\hat{\phi}}_{\tau_i}$ as the AR coefficients of quantile $\tau_i$.
\end{itemize}
Then, the AR coefficients matrix $\hat{\boldsymbol{\Phi}}^*=[\boldsymbol{\hat{\phi}}_{\tau_1},...,\boldsymbol{\hat{\phi}}_{\tau_m}]^{\top}$.

It is possible to obtain $\hat{\boldsymbol{\Phi}}^*$ by smoothing the AR coefficient matrix $\hat{\boldsymbol{\Phi}}$ across the quantile levels, where $\hat{\boldsymbol{\Phi}}$ is obtained by solving the Yule-Walker equation when the orders $\{\hat{p}^*_{\tau_1},...,\hat{p}^*_{\tau_m}\}$ are identified. However, we propose to smooth the quantile PACF matrix $\Psi^*$ in order to ensure they stay between -1 and 1 and therefore satisfy the causal condition requirement that the roots of $g(z)=1-\sum_{k=1}^{\hat{p}^*}\hat{\phi}^*_{\tau_i}(k)z^k$ lie outside the unit circle, where $\hat{\phi}^*_{\tau_i}(k)$ is the $(i,k)$-th element in $\hat{\boldsymbol{\Phi}}^*$. Directly applying a smoother to the AR coefficients does not guarantee this causal requirement.

With the AR coefficients, we have the estimator for $\tau_i$,
\begin{equation}
\hat{f}_{\tau_i}(\omega_l)=\frac{\hat{\sigma}^{*2}_{p_{\tau_i}} }{|g(e^{2\pi j \omega_l})|^2}.
\end{equation}
In cases where the scale parameter is not of primary interest, one may consider the normalized estimator $$ \hat{\tilde{f}}_{\tau_i}(\omega_l)=\frac{\hat{f}_{\tau_i}(\omega_l)}{\sum_{l'} \hat{f}_{\tau_i}(\omega_{l'})},$$ which satisfies $\sum_l \hat{\tilde{f}}_{\tau_i}(\omega_l)=1$. Because of the normalization, the exact value of the scale parameter $\hat{\sigma}^{*2}_{p_{\tau_i}}$ becomes irrelevant. We consider $\hat{\tilde{f}}_{\tau_i}(\omega_l)$ as an estimator of the normalized quantile spectrum $\tilde{f}_{\tau_i}(\omega_l) := \frac{f_{\tau_i}(\omega_l)}{\sum_{l'} f_{\tau_i}(\omega_{l'})}$.

\section{Simulation Studies}\label{q3}
In this section, we describe the simulation setup and then compare the proposed method for estimating the normalized quantile spectrum to three conventional periodogram smoothing methods using two spectral divergence measures.
\subsection{Simulation Setup}
In the simulations, we consider four time series models:
\begin{enumerate}
	\item AR(2) model: $Y_t=0.9Y_{t-1}-0.9Y_{t-2}+w_t$, $w_t\sim N(0,1)$, and $t=1,2,...n$;
	\item ARMA(2,2) model: $Y_t=0.8897Y_{t-1}-0.4858Y_{t-2}+w_t-0.2279w_{t-1}+0.2488w_{t-2}$, $w_t\sim N(0,1)$, and $t=1,2,...n$;
	\item GARCH(1,1) model \citep{engle1982autoregressive}: $Y_t\sim N(0,\sigma_t)$, $\sigma_t^2=10^{-6}+0.35Y^2{t-1}+0.35\sigma_{t-1}^2$, and $t=1,...,n$;
	\item Mixture model: The time series $\{ Y_t \}$ is a nonlinear mixture of three components given by
$$Z_t   :=   W_1(X_{t1}) \, X_{t1} + \{1-W_1(X_{t1})\} \, X_{t2},$$
$$Y_t   :=   W_2(Z_t) \, Z_t + \{1-W_2(Z_t)\} \, X_{t3}.$$
The components $\{X_{t1}\}$, $\{X_{t2}\}$, and $\{X_{t3}\}$ are independent Gaussian AR processes satisfying
$$X_{t1}  = 0.8 X_{t-1,1} + w_{t1},$$
$$X_{t2}  =  -0.75 X_{t-1,2} + w_{t2}, $$
$$X_{t3} =   -0.81 X_{t-2,3} + w_{t3},$$
where $w_{t1}, w_{t2}, w_{t3} \sim N(0,1).$
From the perspective of traditional spectral analysis, the series $\{X_{t1}\}$ has a lowpass spectrum, $\{X_{t2}\}$ has a highpass spectrum, and $\{X_{t3}\}$ has a bandpass spectrum around frequency $1/4$. The mixing function $W_1(x)$ is equal to 0.9 for $x < -0.8$, 0.25 for $x > 0.8$, and linear transition for $x$ in between. The mixing function $W_2(x)$ is similarly defined except that it equals 0.5 for $x < -0.4$ and 1 for $x > 0$.
\end{enumerate}

We generate time series of length $n=(500,1000)$ and we compute the quantile spectrum for each model by averaging 2,000 quantile periodograms of the time series generated from the model, which we treat as the ground truth of the quantile spectrum and show the case in Figure 1 when $n=500$.  The four models have different features. Models 1 and 2 represent a simple and common case with a spectral peak at middle quantiles, respectively, and model 3 is commonly used for financial data such as stock returns with large values at low and high quantiles. Models 1, 2, and 3 have a symmetric quantile spectrum across the quantile levels, while model 4 is asymmetric across the quantile levels. Here, we select 91 quantile levels:  0.05,0.06,...,0.95.

\begin{figure}[!ht]
	\centering
	\subfigure[Quantile spectrum of model 1]{\raisebox{-1cm}{\includegraphics[width=0.45\textwidth]{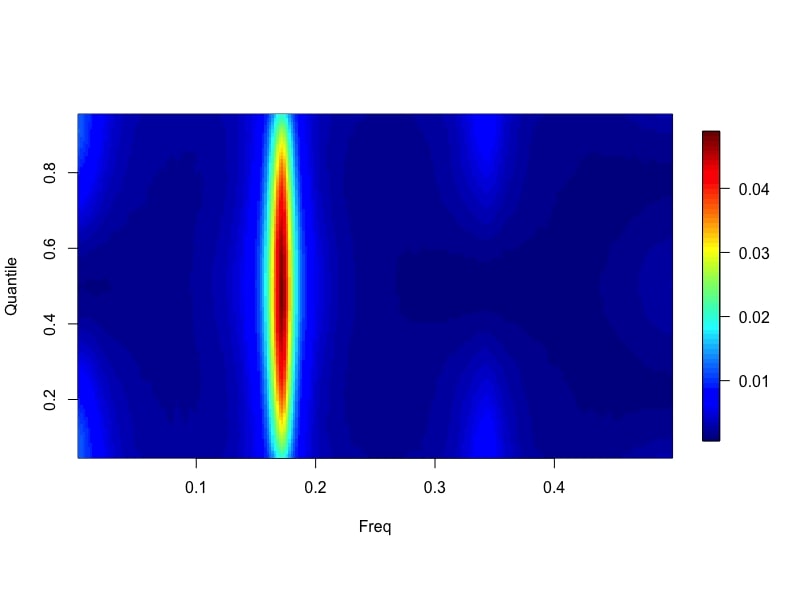}}}
	\subfigure[Quantile spectrum of model 2]{\raisebox{-1cm}{\includegraphics[width=0.45\textwidth]{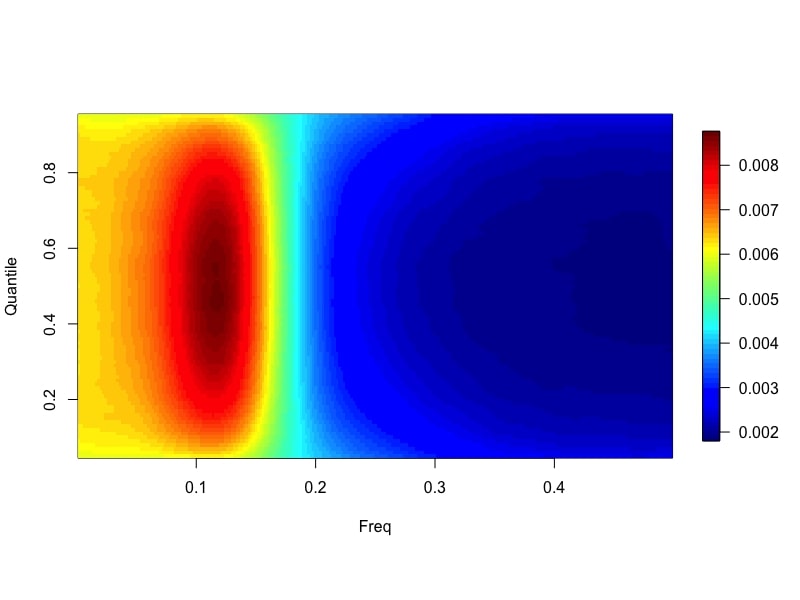}}}
	\subfigure[Quantile spectrum of model 3]{\raisebox{-1cm}{\includegraphics[width=0.45\textwidth]{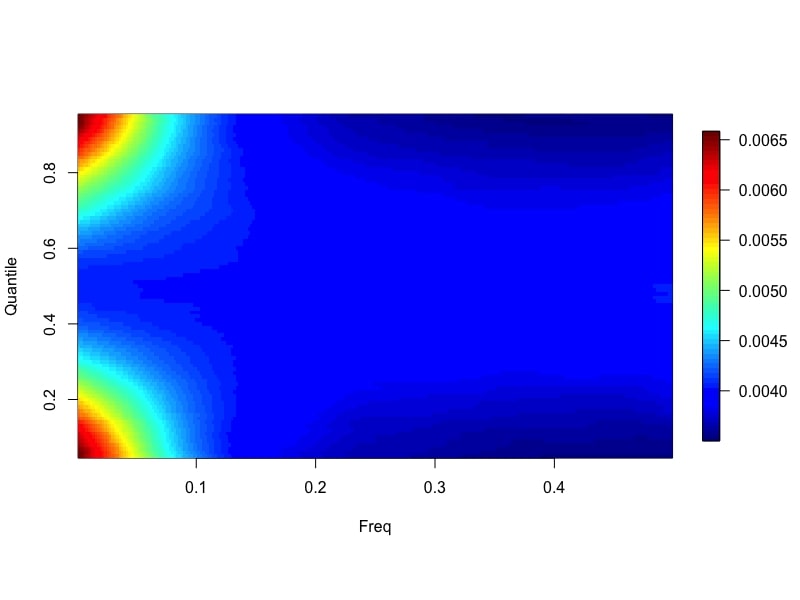}}}
	\subfigure[Quantile spectrum of model 4]{\raisebox{-1cm}{\includegraphics[width=0.45\textwidth]{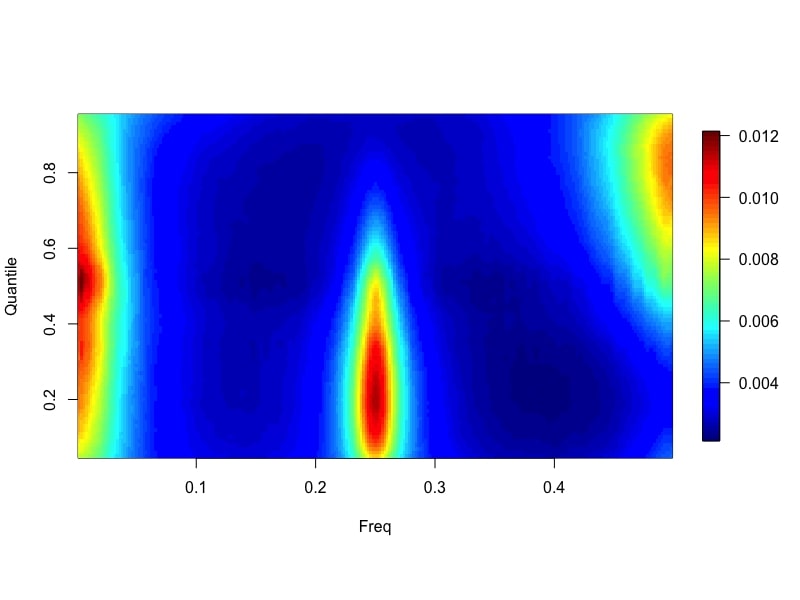}}}
	\label{groundtruth}

	\caption{The true quantile spectra of the four models (n=500). (a) AR(2) model, (b)ARMA(2,2) model, (c) GARCH(1,1) model and (d) mixture model.}
\end{figure}

\subsection{Comparisons}
We compare our proposed algorithm to three commonly used smoothing methods: spline smoothing, Gamma-GCV smoothing \citep{ombao2001simple}, and 2D Gaussian kernel smoothing.
For the spline and Gamma-GCV smoothing, at each quantile level $\tau_i$, $i=1,...,m$, we apply the smoothing methods to the raw quantile  periodograms $Q_{n,\tau_i}(\omega)$, and then smooth across the quantile levels for each frequency using splines. We select the tuning parameter for spline smoothing by ordinary leave-one-out cross-validation (function {\sf smooth.spline} in {\sf R}). We also apply the 2D Gaussian kernel smoothing to the raw quantile periodograms. These three methods, as well as the proposed method, all consider the smoothness of both the frequencies (horizontal) and the quantile levels (vertical).

We use the divergence between the results of each of the estimators and the ground truth to measure the performance. For GARCH models, we only consider the first 1/5 frequencies. For each simulation run of each estimator, we compute the divergence between the estimator and the ground truth at each quantile level $\tau_i$, $i=1,...,m$,  and then average the $m$ distances across the quantile levels. Finally, we use the averaged divergences from the N=200 simulation runs of each estimator to assess performance.

We consider two spectral divergence measures:
\begin{itemize}
	\item  Kullback-Leibler (KL) divergence \citep{kullback1951information}:
	$$D_{KL}\{\tilde{f}_{1,\tau_i}(\omega),\tilde{f}_{2,\tau_i}(\omega)\}=\sum_l \tilde{f}_1(\omega_l){\sf ln}\frac{\tilde{f}_2(\omega_l)}{\tilde{f}_1(\omega_l)}.$$

	\item Root mean squared error (RMSE):
	$$D_{RMSE}\{\tilde{f}_{1,\tau_i}(\omega),\tilde{f}_{2,\tau_i}(\omega)\}=\sqrt{\frac{1}{n}\sum_{l=0}^{n-1}\left[\tilde{f}_{1,\tau_i}(\omega_l)-\tilde{f}_{2,\tau_i}(\omega_l) \right]^2 }.$$
\end{itemize}

\subsection{Simulation Results}
The divergence error from each scenario averaged over N=200 simulations are shown in Tables 1 and 2. The minimum value in each of the four methods in each row is shown in bold. We also include the standard error in the tables.
\begin{table}[!ht]
	\begin{center}
\def\arraystretch{0.5}
		\begin{tabular}{cccccc}
		\hline
Model& n        & Parametric    & Spline      & Gamma-GCV &  2D kernel \\\hline 
AR(2)&500     & {\bf 0.366}(0.151)  & 0.667(0.338)      &0.632(0.165) & 1.606(0.303)\\ 
AR(2)&1000    & {\bf 0.233}(0.085)  & 0.429(0.218)      &0.357(0.088) & 1.643(0.226) \\ 
AR(2)&2000    & {\bf 0.137}(0.045) & 0.240(0.088)      &0.195(0.051)& 1.678(0.160) \\
ARMA(2,2)&500    &{\bf 0.124}(0.039)   &0.156(0.052)    &0.172(0.041) & 1.631(0.142)  \\
ARMA(2,2)&1000   &{\bf 0.074}(0.024)    &0.089(0.028)    &0.164(0.030) & 1.650(0.089)  \\
ARMA(2,2)&2000   &{\bf 0.042}(0.010)   &0.051(0.015)    &0.159(0.021) & 1.686(0.070)  \\
GARCH(1,1)&500   &{\bf 0.030}(0.025)   &0.042(0.036)   &0.111(0.050)& 1.551(0.185)  \\
GARCH(1,1)&1000  &{\bf 0.020}(0.012)   &0.024(0.018)   &0.131(0.046) & 1.633(0.156)   \\
GARCH(1,1)&2000  &{\bf 0.015}(0.008)   & 0.016(0.010)  &0.142(0.031) & 1.679(0.106)  \\
Mixture&500  &{\bf 0.227}(0.062)   &0.257(0.088)       &0.234(0.050) &  1.647(0.124) \\
Mixture&1000  &{\bf 0.139}(0.033)   &0.154(0.040)     &0.185(0.032) &  1.650(0.092)  \\
Mixture&2000  &{\bf 0.082}(0.019)   & 0.092(0.028)    &0.169(0.023) &  1.593(0.060)    \\\hline
		\end{tabular}\\
	\end{center}
	\label{kl}
	\linespread{0.65}
	\caption{The averaged KL divergences to the ground truth  from 200 simulation runs and the standard errors ($\times 10^{-1}$).}
\end{table}

\begin{table}[!ht]
\def\arraystretch{0.5}
	\begin{center}

		\begin{tabular}{cccccc}
		\hline
Model & n      & Parametric    & Spline            &Gamma-GCV  &  2D kernel\\ \hline
AR(2)&500    & {\bf 1.945}(0.461)   & 2.489(0.796)      &2.699(0.401)& 4.124(0.934)   \\
AR(2)&1000   & {\bf 0.806}(0.170)   & 1.046(0.340) &1.054(0.143)& 2.181(0.368)   \\ 
AR(2)&2000  & {\bf 0.313}(0.067)  & 0.399(0.099) &0.352(0.071) &1.132(0.153)   \\
ARMA(2,2)&500  &  {\bf 0.660}(0.124)   &0.741(0.144)        & 0.791(0.115) & 2.751(0.183)   \\
ARMA(2,2)&1000  & {\bf 0.255}(0.046)   &0.281(0.051)        &0.399(0.046) & 1.386(0.061)   \\
ARMA(2,2)&2000  &  {\bf 0.099}(0.013)  & 0.108(0.017)     &0.202(0.017) & 0.703(0.024)   \\
GARCH(1,1)&500   &  0.548(0.134)   &{\bf 0.546}(0.141)    &0.791(0.150)& 2.654(0.247)   \\
GARCH(1,1)&1000  &  0.215(0.051)  &  {\bf 0.213}(0.052)   &0.402(0.063) &  1.364(0.107)\\
GARCH(1,1)&2000  & {\bf 0.082}(0.018)   &{\bf 0.082}(0.019)  &0.201(0.022) &  0.697(0.036)  \\
Mixture&500  &{\bf 0.956}(0.141)   &1.004(0.171)      &0.958(0.118) &  2.622(0.178)  \\
Mixture&1000  &{\bf 0.378}(0.050)   &0.393(0.056)   &0.426(0.048) & 1.355(0.069)   \\
Mixture&2000  &{\bf 0.145}(0.019)   & 0.153(0.024)       &0.205(0.019)&   0.687(0.022) \\\hline
		\end{tabular}\\
	\end{center}
	\label{ks}
	\linespread{0.65}
	\caption{The averaged RMSEs to the ground truth from 200 simulation runs and the standard errors ($\times 10^{-3}$).}
\end{table}

From Tables 1 and 2, we see that the proposed method outperforms the three conventional methods. The proposed method results the smallest divergence to the ground truth under most conditions. There are two exceptions: GARCH model ($n=500,1000$) using RMSE, while the differences are less than $10^{-5}$. Additionally, the divergences and the standard errors show a decreasing trend as $n$ increases.
\section{Earthquake Data Application}\label{q4}

In this section, we apply our method to an earthquake classification problem. We first describe the data in Section \ref{q41} and then present the quantile periodogram analysis in Section \ref{q42}. The classification of the smoothed quantile periodogram using the CNN is illustrated in Section \ref{q43}. 
\subsection{Data Description}\label{q41}
The earthquake data in this section is waveform data collected during February 2014 in Oklahoma (USA). The continuous waveform data and earthquake catalog are publicly available at \\https://www.iris.edu/hq/ and http://www.ou.edu/ogs.html. Details about the catalog data are provided in \cite{benz2015hundreds}, where the magnitudes and times of the earthquakes are labeled. The waveform data we use is from 2014-02-15 00:00:00.005 to 2014-02-28 23:59:58.995, and has a sampling rate of 100 Hz.

\subsection{Quantile Periodogram Analysis}\label{q42}
From the original dataset, there are ten earthquakes with magnitudes greater than 3. Information about the ten earthquakes is shown in Table 3. We extract eight epochs, each comprising one hour of data surrounding one such earthquake event if it is separated from the others by more than one hour or two such earthquake events if they are separated by less than one hour. The total sample size is 2,888,000 (360,000$\times$8). For each epoch, we use a moving window with a width of 20 seconds and move forward at 10 seconds per step, which generates 359 time series of waveform data for each epoch. In each window, the time series is subtracted by a spline fit to eliminate the low-frequency components that do not differ between the earthquake windows and no-earthquake windows. For each time series, we estimate the quantile spectrum using the proposed method and use the first half of the frequencies ($l=1,...,500$) at quantiles \{0.05,0.06,...,0.95\}. The advantage of using a moving window with an overlap is that, for most cases, the earthquakes appear in at least two frames, which makes the visualization easier. The animations that show the evolution of the smoothed periodogram over time for each epoch can be viewed in the supplementary files.
\begin{table}[!ht]
	\begin{center}
\def\arraystretch{0.5}
		\begin{tabular}{|c|c|c|}
			\hline
			Date      & Time    & Magnitude  \\\hline
			02-15-2014  & 00:19:12   & 3.51  \\\hline
			02-16-2014  & 01:51:47   & 3.47  \\\hline
			02-17-2014  & 04:54:59   & 3.50  \\\hline
			02-17-2014  & 05:02:11   & 3.07  \\\hline
			02-17-2014  & 08:24:21   & 3.05  \\\hline
			02-17-2014  & 14:19:13   & 3.36  \\\hline
			02-18-2014  & 11:53:50   & 3.15  \\\hline
			02-18-2014  & 12:16:43   & 3.12  \\\hline
			02-19-2014  & 16:44:56   & 3.25  \\\hline
			02-21-2014  & 21:38:49   & 3.06  \\\hline
		\end{tabular}\\
	\end{center}
	\label{10earth}

	\caption{Information about the ten earthquakes from February 2014.}
\end{table}

Four representative windows from the animations are shown in Figure \ref{frames}, where the window in Figure \ref{frames}(a) contains a very large earthquake with a magnitude of 3.47;  the window in Figure \ref{frames}(b) contains a somehow large earthquake with a magnitude of 1.32; (3) the window in Figure \ref{frames}(c) contains a tiny earthquake with a magnitude $<$ 0.1; and the window in Figure \ref{frames}(d) does not contain an earthquake. By comparing these results, we have the following findings:
\begin{itemize}
	\item The smoothed quantile periodogram in the presence of a large earthquake has a large peak at the low-frequency band in the high and low quantiles (see Figure \ref{frames} (a) and (b)).
	\item The smoothed quantile periodogram without an earthquake has peaks at a higher frequency band (see Figure \ref{frames}(d)).
	\item The smoothed quantile periodogram in the presence of a very small earthquake has peaks at both low frequency band (in low and high quantiles) and high frequency band (in middle quantiles). The magnitude is too small to make the peak at low frequency band overwhelm the peaks at high frequency band (see Figure \ref{frames}(c)).
	\begin{figure}[!ht]
		\centering
		\subfigure[Earthquake window (mag: 3.47)]{\includegraphics[width=0.49\textwidth]{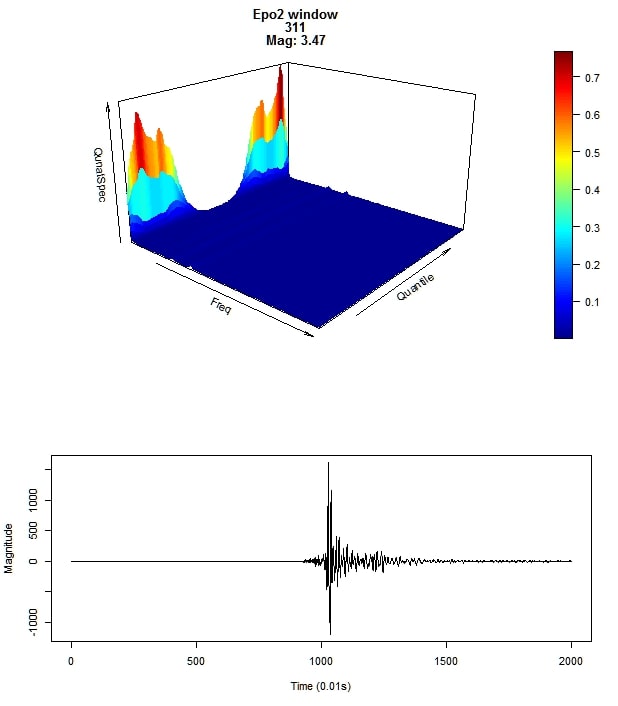}}
		\subfigure[Earthquake window (mag: 1.32
		)]{\includegraphics[width=0.49\textwidth]{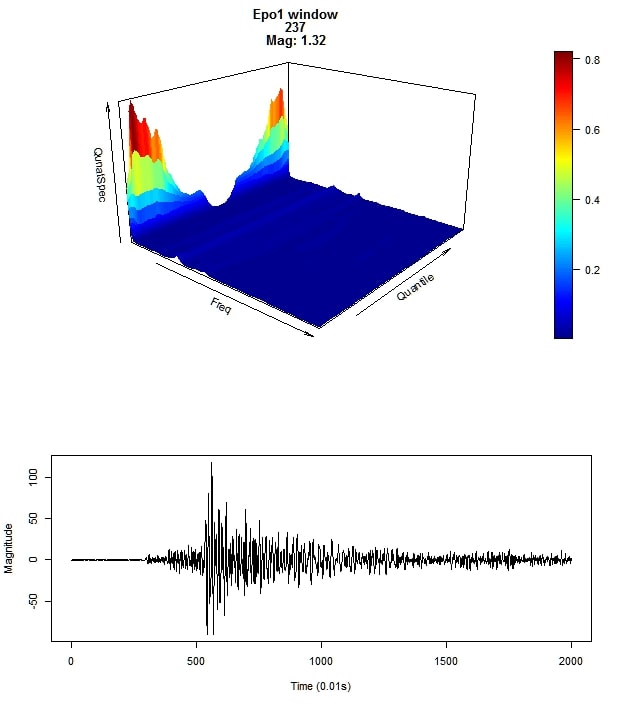}}
		\subfigure[Earthquake window (mag$<$0.1)]{\includegraphics[width=0.49\textwidth]{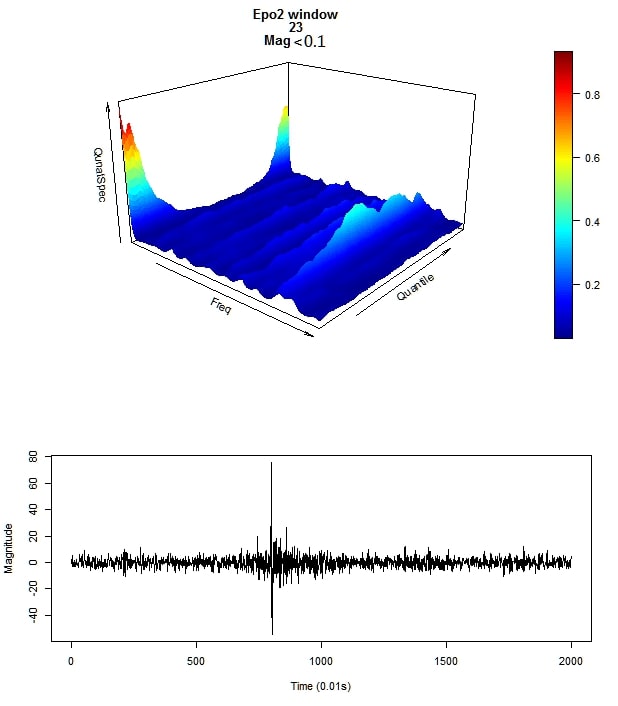}}
		\subfigure[No earthquake window]{\includegraphics[width=0.49\textwidth]{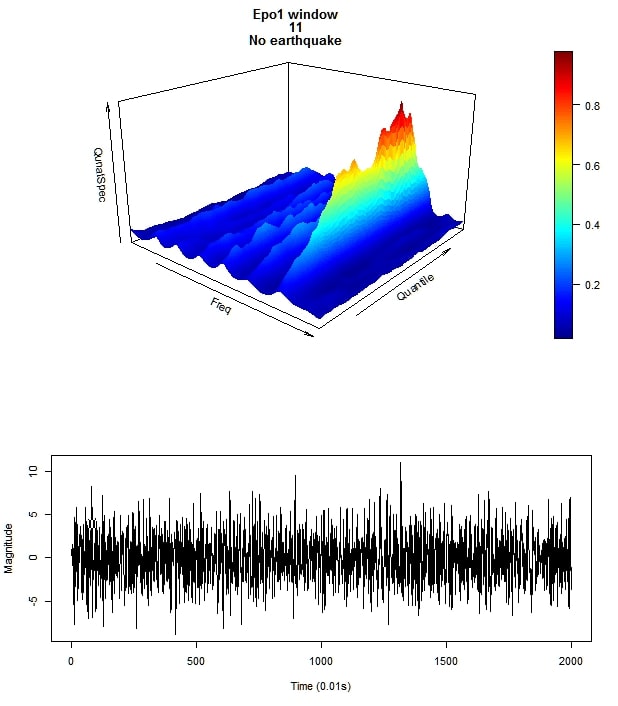}}

\linespread{0.5}
		\caption{(a), (b), and (c) Windows with earthquakes. (d) The window without earthquakes. The top image in each window (a-d) shows a 3D plot of the smoothed quantile periodogram.}
		\label{frames}
	\end{figure}
	
\end{itemize}

\subsection{Classifications Using CNN}\label{q43}

\begin{figure}[H]
	\centering
	\subfigure[CNN structure]{\includegraphics[width=\textwidth]{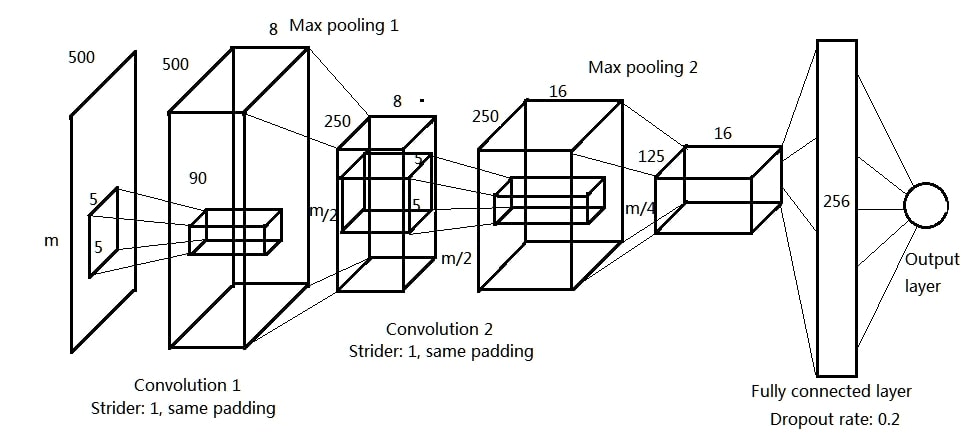}}
	\subfigure[Accuracy curves]{\includegraphics[width=0.54\textwidth]{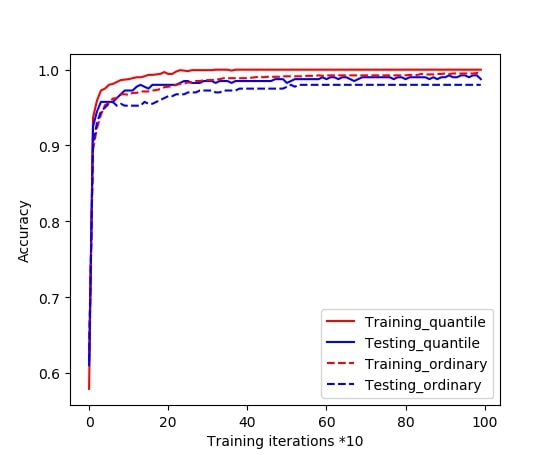}}
	\subfigure[Misclassification  case]{\includegraphics[width=0.45\textwidth]{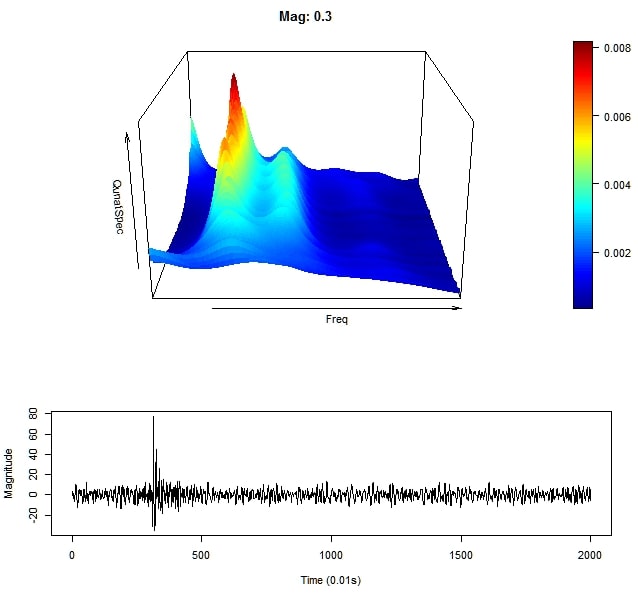}}
	
\linespread{0.5}
	\caption{(a) The CNN structure, (b) the training and testing accuracy of quantile periodograms and ordinary periodograms, and (c) the misclassification case when using quantile periodograms.}
	\label{train}
\end{figure}

In this section, we use the smoothed quantile periodogram as a feature by which we classify the segments into those that contain earthquakes and those that do not. We treat the quantile spectrum as a 2D image and apply the CNN  to build the classifier. We extract 2000 nonoverlapping segments of data, each being a time series of length 2000 (equivalent to 20 seconds). Among the 2000 segments, 1000 of them contain an earthquake with a magnitude higher than 0.25 and the remaining 1000 segments contain no earthquakes. In each segment, the time series is subtracted by a spline fit to eliminate the low-frequency components that do not differ between the earthquake segments and no-earthquake segments. The 2000 smoothed quantile periodogram (we use $l=1,...,500$) and their labels are the input data for the CNN. Since we use the normalized quantile periodograms, the amplitude is not considered;  we focus on the serial dependence, which makes the classifications more challenging. We randomly split the segments into training and testing sets with respective sizes of 1600 and 400. We build the CNN with two convolution layers, each one connected to a maxpooling layer, where the second maxpooling layer connects to fully connected layers (FC). The CNN structure is shown in Figure \ref{train}(a).

To show the advantages of quantile periodograms, we also estimate the ordinary spectra using AR models, with the orders selected by AIC. We apply a CNN with the same structure with different dimensions of the input ($500\times 1$ instead of $500\times m$) and the kernel (5$\times$1 instead of 5$\times$5). We use ten random seeds to initialize the CNN training process and pick the one that yields the best performance. The training and testing accuracy curves are shown in Figure \ref{train}(b), and the confusion matrices are shown in Tables \ref{confusion1} and \ref{confusion2}. The metrics we use are true positive (TP), which indicates that the segment has an earthquake and is classified as an earthquake; true negative (TN), which indicates that the segment has no earthquake and is classified as no earthquake; false positive (FP), which indicates that the segment has no earthquake but is classified as an earthquake; false negative (FN), which indicates that the segment has an earthquake but is classified as no earthquake. From these results, we see that

\begin{itemize}
	\item The classification based on quantile periodograms has a higher accuracy rate. Specifically, for quantile periodograms, the accuracy is 100.00\% (training) and 99.25\% (testing);  for ordinary periodograms, the accuracy is 99.56\% (training) and 98.00\% (testing).
	\item No misclassification in training set and higher precision ($\frac{TP}{TP+FP}$) and recall ($\frac{TP}{TP+FN}$) using quantile periodograms.
	\item One misclassification case using quantile periodograms is shown in Figure \ref{train}(c). Since the magnitude is small, the power at low frequencies is not as large as the power at high frequencies, which cause the misclassification.
\end{itemize}

\begin{table}[H]

	\centering
	\def\arraystretch{0.5}
	\subtable[Training set]{
		\begin{tabular}{|c|c|c|}
			\hline
			&  Positive   & Negative  \\\hline
			True  & 800   & 800  \\\hline
			False  & {\bf 0}   & {\bf 0}  \\\hline
			
		\end{tabular}
	}
	\qquad
	\subtable[Testing set]{
		\begin{tabular}{|c|c|c|}
			\hline
			&  Positive   & Negative  \\\hline
			True& 198 & 199  \\\hline
			False  &  1   & 2  \\\hline
			
		\end{tabular}
	}
	\qquad
	\subtable[Total]{
		\begin{tabular}{|c|c|c|}
			\hline
			&  Positive   & Negative  \\\hline
			True  & 998   & 999  \\\hline
			False  &  1   & 2  \\\hline
		\end{tabular}
	}

	\caption{The confusion matrices of the classification using quantile periodograms.}
	\label{confusion1}
\end{table}

\begin{table}[H]

	\centering
	\def\arraystretch{0.5}
	\subtable[Training set]{
		\begin{tabular}{|c|c|c|}
			\hline
			&  Positive   & Negative  \\\hline
			True  & 795   & 798  \\\hline
			False  & 2   & 5  \\\hline
		\end{tabular}
	}
	\qquad
	\subtable[Testing set]{
		\begin{tabular}{|c|c|c|}
			\hline
			&  Positive   & Negative  \\\hline
			True  & 193   & 199  \\\hline
			False  & 1   & 7  \\\hline
		\end{tabular}
	}
	\qquad
	\subtable[Total]{
		\begin{tabular}{|c|c|c|}
			\hline
			&  Positive   & Negative  \\\hline
			True  & 988   & 997  \\\hline
			False  & 3   & 12  \\\hline
		\end{tabular}
	}
	\caption{The confusion matrices of the classification using ordinary periodograms.}
	\label{confusion2}
\end{table}

\section{Conclusion}\label{q5} 
In this paper, we proposed a new semi-parametric estimation method for the quantile spectrum that uses a parametric form of the AR approximation coupled with nonparametric smoothing. Based on the AR approximation theorem and the maximum entropy principle, we approximate the quantile spectrum by the ordinary AR spectral density at each quantile level. At each quantile level, we define the QACF as the inverse Fourier transform of the raw quantile periodogram. By solving the Yule-Walker equations using the Levinson-Durbin algorithm, we obtain the QPACF and the residual variance from the QACF. The AR order at each quantile is selected by minimizing the AIC, thus balancing the goodness of fit and the model complexity. Then, we smooth the AR orders and the residual variances across the quantile level to reduce variability. Finally, the quantile AR coefficients, obtained from the smoothed QPACF across quantile levels, are used to construct the quantile spectrum. In such a way, the quantile periodogram is smoothed across both the quantile levels and the frequencies. Equipped with the smoothed quantile periodogram, we are able to use the 2D image classification techniques to classify time series. In particular, we apply the CNN, which is a powerful tool for image classification, with the quantile periodogram as input.

The 2D property of the quantile periodogram means that it provides more information than the ordinary periodogram, and that we can take advantage of advanced techniques for images to analyze the time series data. The quantile regression technique also endows the quantile periodogram with sufficient robustness to handle noisy data with outliers. However, the quantile periodogram incurs a higher computational cost because its dimension is multiplied by the number of quantiles. If the scale is not of primary interest, one can use conventional FFT method applied to the clipped data (see \citet{kley2016quantile} and {\sf R} package {\sf quantspec}), which is less efficient statistically but easier to compute, to obtain the level-crossing periodogram instead of using quantile regression to reduce the computational cost. \citet{li2008laplace} and \citet{li2010robust} compared the statistical efficiency of quantile-regression based method and the hard clipping method for hidden periodicity detection and estimation. In the classification procedure using CNN (Section \ref{q43}), the training time was 1.96 seconds per iteration when using quantile periodograms, but only 0.02 seconds per iteration when using ordinary periodograms. One solution to reduce the computational cost is to use fewer quantiles. In this paper, we use a large number of quantiles to make the estimator more like an image. In other applications, we may only look at a few quantiles with sufficient discriminative power, e.g., high or low quantiles. We also used multicore parallelization to speed up computing the raw quantile periodograms.
\section{Acknowledgement}
This research was supported by funding from King Abdullah University of
Science and Technology (KAUST). We would like to thank the editor, associate editor, and reviewers for their valuable comments.\\\\

\bibliography{a1}

\end{document}